\documentclass[sigconf]{acmart}

\usepackage{amsfonts}
\usepackage{amsmath}
\usepackage{graphicx} 
\usepackage{array,color,booktabs,multirow}
\usepackage{balance}
\acmDOI{10.475/123_4}
\acmISBN{123-4567-24-567/08/06}
\acmConference[CIKM'19]{The 28th ACM International Conference on Information and Knowledge Management}{November 2019}{Beijing, China}
\acmYear{2019}
\copyrightyear{2019}
\acmArticle{4}
\acmPrice{15.00}

\begin{document}

\title{MedTruth: A Semi-supervised Approach to Discovering Knowledge Condition Information from \\ Multi-Source Medical Data}
\author{Yang Deng$^{1,2,3}$, Yaliang Li$^4$, Ying Shen$^{1}$, Nan Du$^5$, Wei Fan$^5$, Min Yang$^6$ and Kai Lei$^{1,2,\dagger}$}
\affiliation{$^1$Shenzhen Key Lab for Information Centric Networking \& Blockchain Technology (ICNLAB),\\
  School of Electronics and Computer Engineering, Peking University,\\} 
\affiliation{$^2$PCL Research Center of Networks and Communications, Peng Cheng Laboratory}
\affiliation{$^3$The Chinese University of Hong Kong, $^4$Alibaba Group,\\}
\affiliation{$^5$Tencent Medical AI Lab,$^6$Shenzhen Institutes of Advanced Technology, Chinese Academy of Sciences,\\}
\affiliation{ydeng@se.cuhk.edu.hk, 	yaliang.li@alibaba-inc.com, \{shenying, leik\}@pkusz.edu.cn }
\affiliation{\{ndu, davidwfan\}@tencent.com, min.yang@siat.ac.cn}

\renewcommand{\shortauthors}{Y. Deng et al.}
\renewcommand{\authors}{Yang Deng, Yaliang Li, Ying Shen, Nan Du, Wei Fan, Min Yang, and Kai Lei}

\begin{abstract}
Knowledge Graph (KG) contains entities and the relations between entities. Due to its representation ability, KG has been successfully applied to support many medical/healthcare tasks. However, in the medical domain, knowledge holds under certain conditions. For example, symptom \emph{runny nose} highly indicates the existence of disease \emph{whooping cough} when the patient is a baby rather than the people at other ages. Such conditions for medical knowledge are crucial for decision-making in various medical applications, which is missing in existing medical KGs. In this paper, we aim to discovery medical knowledge conditions from texts to enrich KGs.

Electronic Medical Records (EMRs) are systematized collection of clinical data and contain detailed information about patients, thus EMRs can be a good resource to discover medical knowledge conditions. Unfortunately, the amount of available EMRs is limited due to reasons such as regularization. Meanwhile, a large amount of medical question answering (QA) data is available, which can greatly help the studied task. However, the quality of medical QA data is quite diverse, which may degrade the quality of the discovered medical knowledge conditions. In the light of these challenges, we propose a new truth discovery method, MedTruth, for medical knowledge condition discovery, which incorporates prior source quality information into the source reliability estimation procedure, and also utilizes the knowledge triple information for trustworthy information computation. We conduct series of experiments on real-world medical datasets to demonstrate that the proposed method can discover meaningful and accurate conditions for medical knowledge by leveraging both EMR and QA data. Further, the proposed method is tested on synthetic datasets to validate its effectiveness under various scenarios.
\end{abstract}

\maketitle
\blfootnote{$\dagger$ Corresponding Author} 

\section{Introduction}

Recent years have witnessed many successful real-world applications of Knowledge Graph (KG) in the medical and health domain, such as medical diagnosis \cite{Ni2017Automated,DBLP:conf/icdm/ShenDZLDFYL18}, disease classification \cite{Jiang2017Learning} and drug-drug interaction learning \cite{Yuan2016DrugE,Law2014DrugBank}. A medical KG is composed of medical knowledge triples containing a head entity, a tail entity and the relation between them, and plays an indispensable role in medical knowledge representation and storage.

Despite the success, existing KGs for the medical domain have one missing important component: knowledge triple condition. Due to the uncertainty and complexity of knowledge in the medical domain, knowledge triples are closely linked to some certain conditions, like gender, age and other types of conditions. For example, the knowledge triple \textit{(chest pain, symptom-disease, breast hyperplasia)} should be more likely to be retrieved under the condition of \textit{gender(female)} than \textit{gender(male)}, while the knowledge triple \textit{(cold, disease-drug, little remedies)} is more related to the condition of \textit{age(10)} than \textit{age(70)}, as \textit{little remedies} is a cold medicine for children. Therefore, it is crucial to supplement medical knowledge graph with knowledge condition information.

Electronic Medical Record (EMR) is a structured collection of patient health information and medical knowledge, which contains valuable information about conditions. Thus, it can be a high-quality resource to discover medical knowledge conditions. However, the amount of available EMR data is limited due to many reasons such as regularization and privacy issues. The limited amount of data can result in the loss of important condition information and some inaccurate mined knowledge.

Meanwhile, the rapid emergence of online medical and healthcare Question Answering (QA) communities promotes the communication of medical knowledge and information, and produces a large amount of medical QA data, which can greatly help knowledge condition discovery. But unlike the EMR data, the quality of online medical QA data is hard to guarantee, since the advice or diagnosis provided by website users is based on limited descriptions from patients as well as the professional levels of answers are diverse. If we indiscriminately adopt all the QA data, it may introduce a lot of noise and degrade the quality of discovered conditions.

To tackle the above challenges, we propose a new truth discovery method, MedTruth, for medical knowledge condition discovery, in which the knowledge triples and conditions serve as objects and claims, and each doctor or a user providing answers on QA website is a source. The proposed method has two novel properties: 1) Combining prior source quality information and automatic source reliability estimation; 2) Encoding the object (knowledge triple) information into the proposed method. 

To be more specific, the proposed method first assumes that EMR data is priorly known as high-quality sources, namely reference sources. Current semi-supervised truth discovery methods \cite{Yin2011Semi,Liu2011Online,Dong2012Less} adopt a subset of labeled truth to guide the process of source reliability estimation and truth computation, while the proposed method leverages reference sources to automatically distinguish reliable sources and trustworthy information from unreliable sources and untrustworthy information. Second, the proposed method incorporates the rich contextual information of knowledge triple into truth computation, which enables similar knowledge triples to interactively learn from each other. The interactions between objects have received little attention in existing truth discovery methods \cite{DBLP:journals/pvldb/LinC18,Li2014Resolving,Pasternack2010Knowing}, which regard each object as an independent item without any correlation to other items. We combine the co-occurrence embeddings and the entity embeddings to represent the knowledge triple, and then design a self-attention mechanism to enhance the interaction between similar knowledge triple as well as reduce the interference of noise triples.

To demonstrate the effectiveness of the proposed method, we conduct experiments on both real-world and synthetic datasets. We first evaluate the performance of the proposed method on the medical knowledge condition discovery task with a real-world medical dataset, containing $41,700$ EMRs from $360$ doctors and $275,262$ QA pairs answered by $12,501$ users. The experimental results show that the proposed method can overcome the lack of high-quality medical data and effectively discover accurate medical knowledge conditions.
Further, we validate the performance of the proposed method under various scenarios including single truth finding and multiple claims ranking tasks. We also conduct experiments to verify the effectiveness of two novel properties of the proposed method by visualizing how reference sources semi-supervise the other sources in source reliability estimation and how object embeddings enhance the interactions between objects in truth computation.

To summarize, the main contributions of this paper are as follows:

\begin{itemize}
\item We enrich medical knowledge graph with condition information by discovering the knowledge triple condition information from multi-source medical textual data, which enables medical KG to be more accurate and more applicable for medical tasks.
\item We design a novel truth discovery method, MedTruth, for medical knowledge condition discovery, which employs high-quality sources to semi-supervise truth discovery task and incorporates object information to capture the interaction between objects in the process of information trustworthiness estimation.
\item We validate the proposed method on both real-world and synthetic datasets. On real-world medical datasets, we show that the proposed method can reliably discover knowledge triple condition for medical knowledge graph. On synthetic datasets, we demonstrate the effectiveness of the proposed method under various scenarios.
\end{itemize}

\section{PROBLEM DEFINITION}
In this section, we first introduce some important concepts of input and output, and then formally define the task. 

\begin{table}
\setlength{\abovecaptionskip}{0pt}   
\setlength{\belowcaptionskip}{0pt}
\fontsize{7}{9}\selectfont
\centering
  \caption{An Example of EMR Data}
  \begin{tabular}{cp{6cm}}
    \toprule
	 Item & Content\\
     \midrule
     GENDER & Female \\
     AGE & 18\\
     \multirow{2}{*}{ILLNESS\_DESC} & The patient experienced nausea, vomiting, abdominal pain, diarrhea after eating discomfort.\\
     BODY\_EXAM & Left upper abdomen and left lower abdomen have tenderness.\\
     DIAG\_DESC & Acute gastroenteritis.\\
     $\cdots$&$\cdots$\\
    \bottomrule
\label{emrexp}
\vspace{-0.6cm}
\end{tabular}
\end{table}

\begin{table} 
\setlength{\abovecaptionskip}{0pt}   
\setlength{\belowcaptionskip}{0pt}
\fontsize{7}{9}\selectfont
\centering
  \caption{An Example of QA Data}
  \begin{tabular}{cp{6cm}}
    \toprule
	 Item & Content\\
     \midrule
     GENDER & Male \\
     AGE & 66\\
     \multirow{2}{*}{QUESTION} & I got hemoptysis, cough, chest pain, general malaise, trembling. What caused these symptoms?\\
     \multirow{2}{*}{ANSWER} & According to your situation, consider the bronchiolitis or bronchitis that causes capillary bleeding.\\
     $\cdots$&$\cdots$\\
    \bottomrule
\label{qaexp}
\vspace{-0.7cm}
\end{tabular}
\end{table}

\subsection{Input}

The inputs of this task are a set of electronic medical records (EMRs) and a set of question-answer (QA) pairs. The examples of EMR and QA data are shown in Table~\ref{emrexp} and~\ref{qaexp}, respectively.

\textbf{Definition 1}: A \textit{Knowledge Triple} is composed by (head entity, relation, tail entity). A \textit{Knowledge Triple Mention} is a knowledge triple extracted from the text of an EMR or a QA pair.

\textbf{Definition 2}: The \textit{Condition} of a knowledge can be gender, age, onset season and other types of conditions. 

\textbf{Definition 3}: Each EMR or QA pair is a \textit{case} that contains several knowledge triples under certain conditions.

\textbf{Example 1}: Given the \textit{case} from QA data in Table~\ref{qaexp}, we extract the knowledge triple mentions \textit{(chest pain, symptom-disease, bronchiolitis)}, \textit{(chest pain, symptom-disease, bronchitis)} and the others as well as their corresponding conditions \textit{gender(male)} and \textit{age(66)}.

\textbf{Definition 4}: A doctor providing EMRs or a user answering online questions is considered as a \textit{source}. The sources in EMR data are priorly known as high-quality sources and regarded as \textit{reference sources}, while the sources in QA data are regarded as \textit{non-reference sources}. 

\textbf{Definition 5}: A condition is considered as a \textit{claim}. A knowledge triple is considered as an \textit{object}, subject to some certain conditions given by sources. 

\textbf{Example 2}: As the \textit{case} in Example 1, \textit{(chest pain, symptom-disease, bronchiolitis)} and \textit{(chest pain, symptom-disease, bronchitis)} are two objects that are subject to two claims \textit{gender(male)} and \textit{age(66)}; this case is provided by a \textit{non-reference source} from QA data.

\subsection{Output}

\textbf{Definition 6}: \textit{Source Reliability Weight} $\{\omega_k\}$ measures the reliability degree of sources. The higher the $\omega_k$, the more reliable the $k$-th source.

\textbf{Definition 7}: \textit{Truth Vector} $\{v_m^*\}$ is the embeddings of the most-likely correct condition for the knowledge triple $f_m$.

\textbf{Definition 8}: A \textit{Conditional Knowledge Graph} is a set of triples $(f_m,c_n,p_{(m,n)})$, where $c_n$ is the given condition and $p_{(m,n)}$ is the confidence score of $c_n$ for $f_m$. The higher the $p_{(m,n)}$, the more relevant the $f_m$ to the $c_n$.

\subsection{Task Definition}

Based on the definitions, we can formulate the medical knowledge condition discovery task as follows: Given an EMR set and a QA pair set, the goal is to compute the confidence score $\{p_{(m,n)}\}$ of certain conditions $\{c_n\}$ for KG triples $\{f_m\}$ that are mentioned in the EMR and QA data. 

First, we extract triple mentions and corresponding conditions from the text of both EMRs and QA pairs. Then, we design a truth discovery method to automatically estimate the source reliability $\{\omega_k\}$ and compute the truth vector $\{v_m^*\}$. Finally, we calculate the confidence score $\{p_{(m,n)}\}$ of discovered conditions for knowledge triple mentions. 

Table~\ref{notation} summarizes the notations used in this paper. For some notations, more detailed explanations will be introduced in the following section.

\begin{table}
\setlength{\abovecaptionskip}{0pt}   
\setlength{\belowcaptionskip}{0pt}
\fontsize{7}{9}\selectfont
\centering
  \caption{Notations.}
  \begin{tabular}{cp{7cm}}
    \toprule
	Notation & Definition  \\
    \midrule
    $f_m$ & $m$-th knowledge triple\\
    $c_n$ & $n$-th condition\\
    $p_{(m,n)}$ & the confidence score of $c_n$ for $f_m$\\
    $v_m^*$ & the truth vector for $f_m$\\
    $u_n$ & vector representation of $c_n$\\
    $\omega_k$ & reliability degree of $k$-th source\\
    $F_k$ & set of tuples (knowledge triple, condition) that $k$-th source provides \\
    \multirow{2}{*}{$F_{ref}$} & set of tuples (knowledge triple, condition) that the reference sources provide \\
    $C_m^n$ & set of knowledge triples that occur with $f_m$ and $c_n$ in the same case\\
  \bottomrule
\label{notation}
\vspace{-0.5cm}
\end{tabular}
\end{table}

\section{METHODOLOGY}
In this section, we introduce the proposed method, MedTruth, for medical knowledge condition discovery task, which leverages EMR data and QA data to enrich the KG with knowledge triple condition information.

Under the circumstance that the amount of high-quality medical data is limited and there is a large amount of crowdsourcing medical data with diverse quality, we aim to exploit the priorly known reliable sources to semi-supervise the overall truth discovery process. As mentioned in Section I, the interaction between objects should be taken into account in the medical knowledge condition discovery task. To be specific, two similar knowledge triples should have similar condition information. In order to measure the similarity between knowledge triples, we consider the truth discovery procedure as a representational learning procedure to learn a truth vector $v_m^*$ for each knowledge triple $f_m$. Correspondingly, we also learn a representation vector $u_n$ for each condition $c_n$, so that if two knowledge triples are close in representation space, their distance to given conditions will be similar. 

The general principle of truth discovery and the aforementioned motivations are formulated by the following objective function: 
\begin{small}
\begin{equation}
\begin{split}
 \min_{\{\omega_k\},\{v_m^*\},
 \{u_n\}} &\sum_{k=1}^K \frac{\omega_k}{|F_k|+\lambda|\Delta_k|} \sum_{(m,n)\in F_k} 
(1+\lambda^*) \Bigg\{  \\
& \left(v^*_m - u_n\right)^2 + \mu \sum_{i\in {C_m^n}} \alpha_i \left(v_m^* - x_i\right)^2 \Bigg\}, \\
 s.t.&\quad \sigma\left(\omega_k\right) = \sum_{k=1}^K \exp \left(-\omega_k\right) = 1.\\
\end{split}
\end{equation}
\end{small}where $\Delta_k = F_k \cap F_{ref}$, 
$\lambda^* =  \left\{
 \begin{array}{ll}
 \lambda, \quad (m,n)\in \Delta_k \\
 0,\quad (m,n) \not\in \Delta_k\\
 \end{array}  
 \right.$.

Besides the notations in Table~\ref{notation}, there are some notations in the objective function remained to be introduced. $\Delta_k$ contains the tuple (knowledge triple, condition) that occurs both in $k$-th source and the reference sources. $\lambda$ is a parameter to measure the semi-supervised degree of reference sources. $\sigma(*)$ is a regularization function on the source reliability. $x_i$ is the knowledge triple embedding constructed by:
\begin{equation}
x_i = [t_i:\frac{1}{2}(e_h+e_t)],
\end{equation}
where $t_i$ is learned by CBOW model \cite{Mikolov2013Efficient}. The knowledge triple mentions in the same case, regarded as the words in the same sentence, are input into the CBOW model to learn the co-occurrence embeddings of the knowledge triple. $e_h$ and $e_t$ are the pre-trained entity embeddings of head entity and tail entity in the knowledge triple. $\mu$ is used to adjust the importance of the knowledge triple embeddings. $\alpha_i$ is the attention weight over the knowledge triples in the same case, calculated by:
\begin{equation}
\alpha_i = \frac{\exp\left({x_i}^T x_m\right)}{\sum_{j\in {C_m^n}}\exp\left({x_j}^T x_m \right)},
\end{equation}
where $x_m$ is the embeddings of current knowledge triple $f_m$.

The basic ideas behind the objective function are summarized as follows: 

\begin{itemize}
\item The first term, $\left(v^*_m - u_n\right)^2$, in the objective function minimizes the weighted deviation between the representation of the given condition and the truth vector for the knowledge triple. If a source (i.e., doctor or user) has a high-reliability degree, the representations of conditions provided by this source should be close to the truth vectors for the corresponding knowledge triples;

\item Vice versa, if a source provides conditions that are close to the truth for the corresponding knowledge triple, this source should be assigned a high reliability. The first two points are in accordance with the general principle of truth discovery; 

\item Following the motivation that the information provided in the reference source is more trustworthy, the knowledge triples are assumed to be more related to the given condition provided in the reference source. The parameter $\lambda$ increases the weight of information in the tuples (knowledge triple, condition) that occur in reference sources, since this information would be more valuable and explanatory in estimating the source reliability;

\item The second term, $\sum_{i\in {C_m^n}} \alpha_i \left(v_m^* - x_i\right)^2$, is designed to incorporate knowledge triple embeddings to measure the correlation between knowledge triples. 
Two knowledge triples that share similar knowledge or co-occurrence information should also share similar condition information.
If a source has a high-reliability degree, knowledge triples in the source should have a greater influence on their similar knowledge triples ---- that is to say, the truth vectors for knowledge triples in the source should be close to their embeddings.

\item Another intuition behind the second term is that the contextual information for knowledge triples should not be abandoned. Thus, we exploit all the triples in the same case to represent the current triple. However, this raises a new issue that not all the triples contribute equally to the representation of the current triple. To tackle this issue, we employ a self-attention mechanism to weight the triples by their embedding similarity with the current triple. There are two major advantages of the attention mechanism: 1) Enhance the interaction between similar triples and reduce the interference of unrelated triples; 2) Normalize the second term, even if the number of triples is various in different cases.

\end{itemize}

In the proposed objective function, there are three sets of variables need to be solved: truth vectors for knowledge triples $\{v_m^*\}$, vector representations for conditions $\{u_n\}$ and source reliability $\{\omega_k\}$. By solving the objective function in Eqn. (1), it leads to an iterative procedure, in which truth vector learning step, condition representation learning step and source reliability estimation step are iteratively conducted until convergence.

\subsection{Truth Vector Computation}

In the truth vector learning step, condition vector representations $\{u_n\}$ and source reliability $\{\omega_k\}$ are assumed to be fixed. Then the truth vectors for knowledge triples $\{v_m^*\}$ can be inferred by solving the following optimization problem:
\begin{small}
\begin{equation}
\begin{split}
 \min_{\{v_m^*\}} \sum_{m=1}^M \sum_{k\in K_m} \sum_{n\in N_m^k}&\frac{\omega_k (1+\lambda^*)}{|F_k|+\lambda|\Delta_k|}  
 \Bigg\{\left(v^*_m -u_n\right)^2 \\
&  + \mu \sum_{i\in C_m^n} \alpha_i \left(v_m^* - x_i\right)^2 \Bigg\} ,
\end{split}
\end{equation}
\end{small}where $K_m$ is the set of sources that provides $f_m$, and $N_m^k$ is the set of conditions that $f_m$ is subject to in $k$-th source.

The above optimization problem can be split into $M$ separate optimization problems for each knowledge triple. The truth vector for each knowledge triple is computed by the weighted mean of vector representations of related conditions and attentive embeddings of knowledge triples in the same context: 
\begin{small}
\begin{equation}
 v_m^* = \frac{\sum_{k\in K_m}  \sum_{n\in N_m^k} \frac{\omega_k (1+\lambda^*)}{|F_k|+\lambda|\Delta_k|} \left( u_n  + \mu\sum_{i\in C_m^n}\alpha_i x_i\right)}{\sum_{k\in K_m} \sum_{n\in N_m^k}\frac{\omega_k\left(1+\lambda^*\right)}{|F_k|+\lambda|\Delta_k|} \left(1+\mu\right)}.
\end{equation}
\end{small}
We further analyze Eqn. (5) with the following explanations:

\begin{itemize}
\item The information provided by reliable sources is more valuable for computing the truth vector of the same knowledge triple, including both the condition information and the contextual information. In other words, the truth vector $v_m^*$ is close to the condition vector $u_n$ and the contextual embeddings $\sum_{i\in C_m^n}\alpha_i x_i$ provided by the reliable sources.

\item Due to the effect of $\lambda$, averagely speaking, the truth vector $v_m^*$ will be closer to the condition vector $u_n$ and the contextual embeddings $\sum_{i\in C_m^n}\alpha_i x_i$ in the reference sources than the non-reference sources;

\item Two knowledge triples that share similar embeddings will obtain similar truth vectors $v_m^*$ so that they will learn similar condition information, which enables infrequent triples to learn from their similar triples. Besides, two triples provided by the same case $C_m^n$ will have relatively similar representation $v_m^*$ and share similar condition information.
\end{itemize}

\subsection{Condition Representation Learning}

In the condition representation learning step, the truth vectors for knowledge triples $\{v_m^*\}$ and source reliability $\{\omega_k\}$ are assumed to be fixed. Then condition vector representations $\{u_n\}$ can be inferred by solving the following optimization problem:
\begin{small}
\begin{equation}
\begin{split}
 \min_{\{u_n\}}  \sum_{n=1}^N\sum_{k=1}^K  \sum_{m\in M_n^k} \frac{\omega_k (1+\lambda^*)}{|F_k|+\lambda|\Delta_k|}  
 \left\{\left(v^*_m - u_n\right)^2 + b_m \right\} ,
\end{split}
\end{equation}
\end{small}where $M_n^k$ is the set of knowledge triples that are subject to $c_n$ in $k$-th source. $b_m=\mu \sum_{i\in {C_m^n}} \alpha_i \left(v_m^* - x_i\right)^2$ can be regarded as a constant, since it is irrelevant to $u_n$. The above optimization function can be split into $N$ separate optimization problems for each condition. The representation vector for each condition is computed by the weighted mean of truth vectors for corresponding knowledge triples: 
\begin{equation}
\begin{split}
 u_n = \frac{\sum_{k=1}^K  \sum_{m\in M_n^k}\frac{\omega_k}{|F_k|+\lambda|\Delta_k|} (1+\lambda^*) v_m^*}{\sum_{k=1}^K \sum_{m\in M_n^k} \frac{\omega_k}{|F_k|+\lambda|\Delta_k|}(1+\lambda^*)}.
\end{split}
\end{equation}

We have the following intuitions in Eqn. (7):
\begin{itemize}
\item The condition representation vector $u_n$ is close to the truth vector $v_m^*$ for the corresponding knowledge triple provided by the source with a high reliability weight;

\item The condition representation vector $u_n$ tends to be closer to truth vectors $v_m^*$ for the corresponding knowledge triples in reference sources.
\end{itemize}

\subsection{Source Reliability Estimation}

In the source reliability estimation step, source reliability weights $\{\omega_k\}$ are estimated based on the current inferred truth vectors $\{v_m^*\}$ and condition vectors $\{u_n\}$, by solving the following optimization problem:
\begin{small}
\begin{equation}
\min_{\{\omega_k\}} \sum_{k=1}^K \omega_k \cdot \theta_k, \quad 
 s.t.\quad \sigma\left(\omega_k\right) = \sum_{k=1}^K \exp \left(-\omega_k\right)  = 1.
\end{equation}
\end{small}
where the weighted mean of errors that $k$-th source produces can be regarded as a constant $\theta_k$:
\begin{small}
\begin{equation}
\theta_k\! =  \!\frac{1}{|F_k|\!+\!\lambda|\Delta_k|} \!\sum_{(m,n)\in F_k} \!
(1\!+\!\lambda^*)\! \left\{\!\left(v^*_m\! -\! u_n\right)^2\! +\! \mu\! \sum_{i\in C_m^n} \!\alpha_i \!\left(v_m^*\! -\! x_i\right)^2 \!\right\}.
\end{equation}
\end{small}

Thus, the optimization problem can be solved by:
\begin{equation}
 \omega_k = -\log \left(\frac{\theta_k}{\sum_{k=1}^K \theta_k}\right).
\end{equation}

The intuitions behind the source reliability estimation step can be summarized as follows:
\begin{itemize}
\item Following the general truth discovery principle, the proposed method assigns a high reliability degree to the source that provides conditions that are close to the truths of the corresponding knowledge triples;

\item The distance between the truth vector $v_m^*$ and its contextual embeddings $x_i$ also affects the source reliability degree. 
As discussed above, this allows the proposed method to exploit the knowledge triple embeddings to capture the interaction between objects;

\item The errors terms about the tuple (knowledge triple, condition) that appears in the reference sources make more contribution to the reliability degree estimation of current source. As for the reference sources, due to the weighted mean term $\frac{1}{|F_k|+\!\lambda|\Delta_k|}$, $\lambda$ will not affect the source reliability. 
But for the non-reference sources, due to the effect of $\lambda$, the formula (9) will assign more weight on the error of tuples (knowledge triple, condition) that appear in the reference sources. This prevents the reference sources to dominate the overall truth discovery process and also encodes the intuition that reference sources semi-supervise the source reliability estimation procedure.
\end{itemize}

\begin{figure*}[htb]
\setlength{\abovecaptionskip}{0pt}   
\setlength{\belowcaptionskip}{0pt}
\centering
\includegraphics[width=0.9\textwidth]{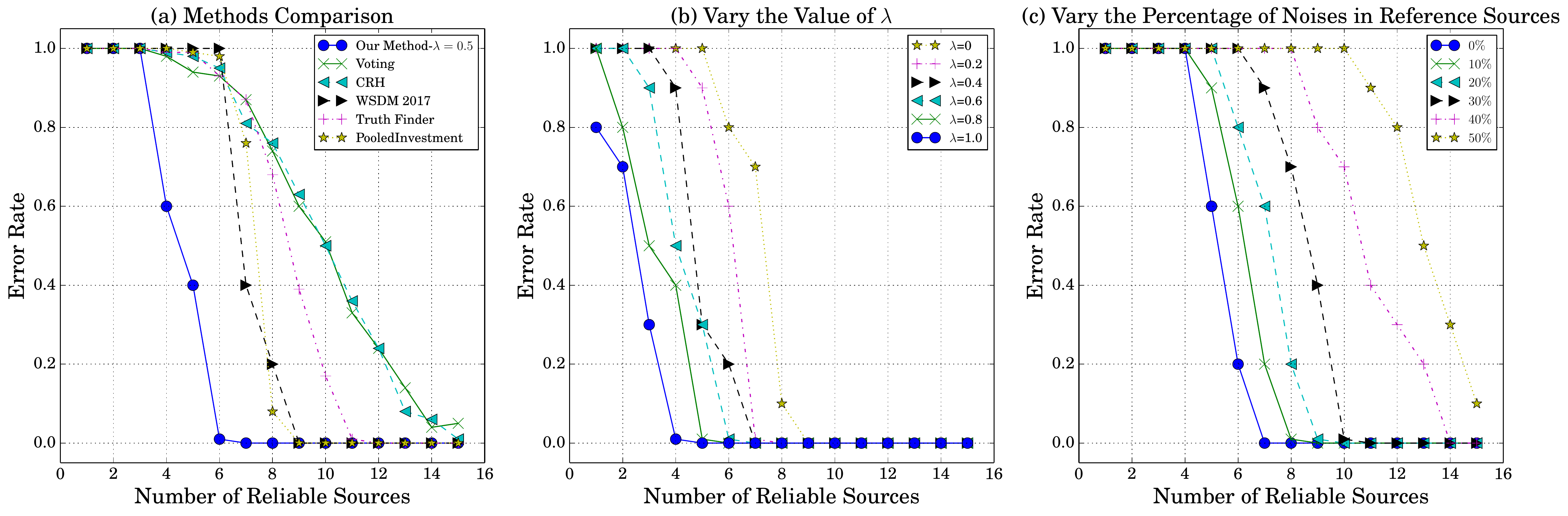}
\caption{Single Truth Finding}
\label{exp_5A}
\vspace{-0.3cm}
\end{figure*}

\subsection{Initialization and Post-processing}

\subsubsection{Source Reliability Initialization}
We adopt the similarity between reference sources and non-reference sources to initialize the source reliability. The similarity is measured by calculating the percentage of common tuples (knowledge triple, condition) provided by both the current source and reference sources: 
\begin{equation}
\omega_k^{(init)} = \frac{|F_k\cap F_{ref}|}{|F_k|+|F_{ref}|}.
\end{equation}

\subsubsection{Truth Vector Initialization}
According to Eqn. (5), we initialize the truth vectors $\{v_m^*\}$ only based on the contextual embeddings for the knowledge triples by:
\begin{equation}
 v_m^* = \frac{\sum_{k\in K_m}  \sum_{n\in N_m^k} \frac{\omega_k (1+\lambda^*)}{|F_k|+\lambda|\Delta_k|} \sum_{i\in C_m^n}\alpha_i x_i}{\sum_{k\in K_m} \sum_{n\in N_m^k}\frac{\omega_k\left(1+\lambda^*\right)}{|F_k|+\lambda|\Delta_k|}}.
\end{equation}

\subsubsection{Condition Confidence Calculation}
The truth vectors $\{v_m^*\}$ for knowledge triples and the condition vector representations $\{u_n\}$ can be obtained by the above truth discovery procedure. The Euclidean Distance between the condition and the truth vector is adopted to calculate the confidence score of the condition for the corresponding knowledge triple as:
\begin{equation}
 p_{(m,n)} = \frac{\min_{j}(v_m^* - u_j) ^2}{(v_m^* - u_n) ^2}.
\end{equation}

\section{Experiment on Synthetic Data}
In this section, we conduct a series of experiments on synthetic datasets to demonstrate the effectiveness of the proposed method. 

\subsection{Single Truth Finding}
We construct a source with 100 different objects and 10 different claims as the ground truth, where each object has a corresponding truth claim. We generate reference sources and non-reference sources by randomly changing claims from the ground truths with 5\% and 95\% noise, respectively. In order to demonstrate the effectiveness of the proposed method under the situation in which there are limited reliable sources, we fix the total number of sources as 100, and increase the number of reference sources from 1 to 15 to evaluate the error rate of single truth finding.

\subsubsection{Method Comparison} We compare the proposed method with several widely-used truth discovery methods as follows. First of all, the majority voting approach is the most widely adopted baseline for truth discovery, which takes the majority claims as truths. TruthFinder \cite{Yin2007Truth} is proposed based on Bayesian analysis, in which the probability of a claim is computed with an assumption that each source's reliability is the probability of it being correct and the source reliability score is calculated by averaging the probability of claims. Investment and PooledInvestment \cite{Pasternack2010Knowing} both assume that a source ``invests" its reliability on the claims it provides, and the trustworthiness of each claim is computed by a non-linear function while the reliability of a source is generated by the weighted sum of the trustworthiness of claims it provides. CRH \cite{Li2014Resolving} is a truth discovery framework that can resolve conflicts in heterogeneous data including categorical data and continuous data. The truth discovery method in \cite{li2017reliable} (we name it as \emph{WSDM 2017}) aims to capture the semantic meanings of claims and combine the truth discovery and vector learning processes. For each baseline, we follow the experimental and parameter settings as the original paper. 

Fig.~\ref{exp_5A}(a) summarizes the results of all the methods on the synthetic dataset. In general, the results show that the proposed method outperforms all the baselines with various number of reliable sources. The proposed method identifies all the truths even only 6 out of 100 sources are reliable, which indicates that the proposed method can effectively infer the truths in the situation where there are only a small amount of reliable sources.

\subsubsection{Vary the Size of $\lambda$} In the proposed method, the parameter $\lambda$ measures the semi-supervised degree of the reference sources. In order to study the effect of $\lambda$ on the semi-supervised truth discovery, we vary the value of $\lambda$ from 0 to 1.0 with 6 different values. The results presented in Fig.~\ref{exp_5A}(b) indicates that if the reference sources are reliable, the higher the semi-supervised power of the reference sources, the less the reference sources are needed.

\subsubsection{Vary the Percentage of Noise in Reference Sources} Although we assume the reference sources are more reliable than the other sources, the actual reliability of reference sources will still influence the truth finding results. Thus, we vary the reliability of reference sources by increasing the percentage of noise in them to study the impact of the actual reference source reliability on semi-supervised truth finding. We can observe from Fig.~\ref{exp_5A}(c) that given the same supervised degree, the error rate increases with the increasing noise percentage in reference sources.

\subsection{Multiple Claims Ranking}
We construct a source with 100 different objects and each object has several ranked related claims as the ground truth. The rank of related claims is determined by the similarity of assigned truth vectors and claim vectors. Then the object embeddings are obtained by inverse operation of the proposed method. The reference sources and non-reference sources are generated in the same way as that in Single Truth Finding (Section 4.1). The experimental settings are also in accordance with Single Truth Finding, but Mean Reciprocal Rank (MRR) and Mean Average Precision (MAP) are adopted to evaluate the results of ranking related claims.

\begin{figure}
\setlength{\abovecaptionskip}{0pt}   
\setlength{\belowcaptionskip}{0pt}
\centering
\includegraphics[width=0.5\textwidth]{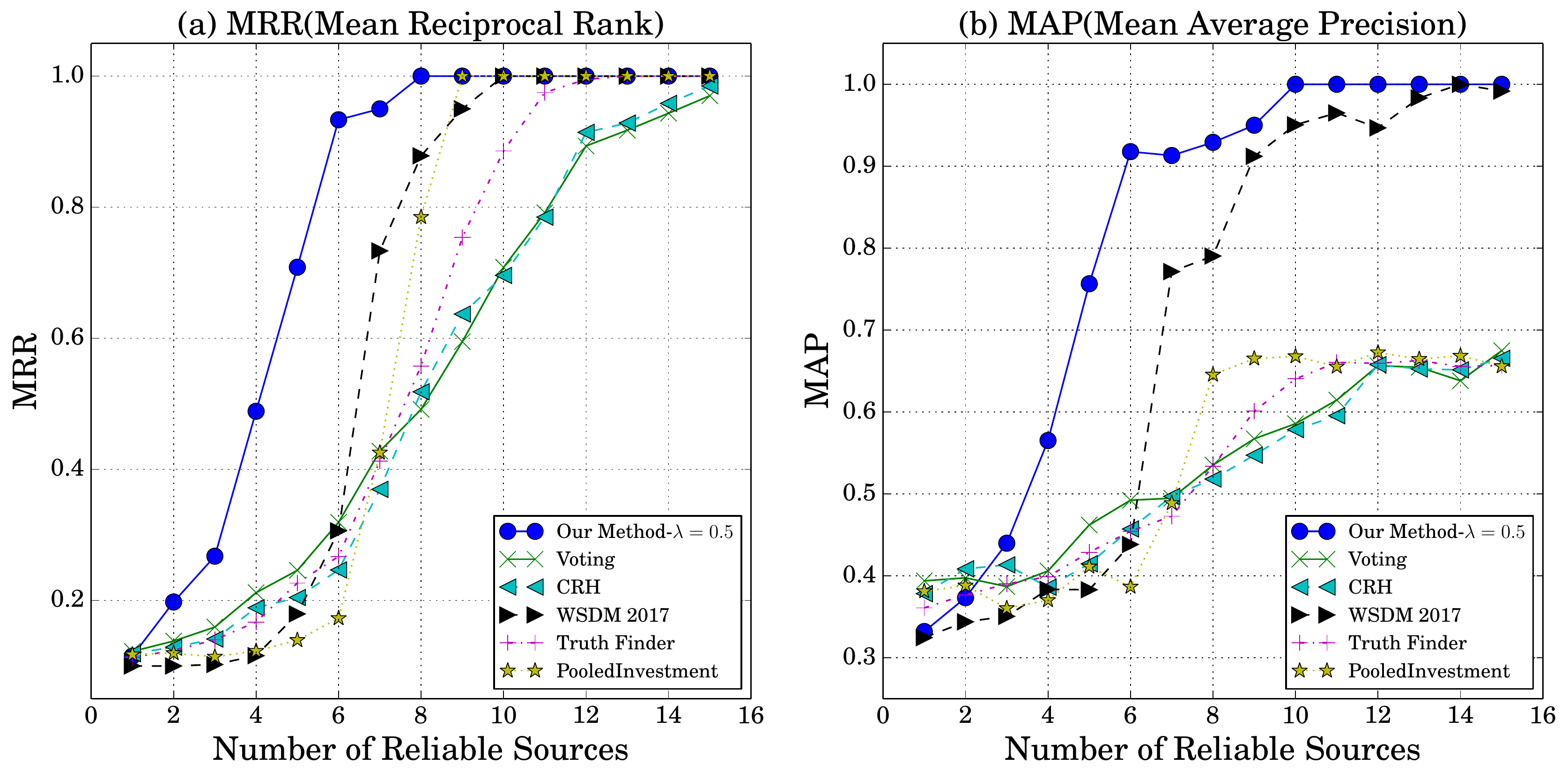}
\caption{Multiple Claims Ranking}
\label{exp_5B}
\vspace{-0.3cm}
\end{figure}

As is shown in Fig.~\ref{exp_5B}, we observe that the proposed method substantially outperforms the existing methods by a noticeable margin on both MRR and MAP matrices. The following observations can be made from these results: (1) In general, the result in Fig.~\ref{exp_5B}(a) shows that it is easier to detect the most related claim when there are a larger amount of reliable sources. However, even when only 8 out of 100 sources are reliable, the proposed method can still discover the most related claim, while the existing methods need more reliable sources to achieve similar performance. This result indicates that the proposed method can effectively infer the most related claim even with limited reliable sources. (2) In Fig.~\ref{exp_5B}(b), only the proposed method and WSDM 2017 achieve 1.0 on MAP, which indicates that these methods can discover all the related claims. Since the proposed method and WSDM 2017 take into account the correlation between objects or claims, even if a related claim is not provided by any source for the corresponding object, these two methods can learn from other claims or objects with similar embeddings. Compared with WSDM 2017, the proposed approach can successfully rank all the related claims with less reliable sources. On the contrary, for other methods, object and claim are independent with each other. Therefore, if a related claim is not provided by any source for the corresponding object, the claim will be irrelevant to the object when using these methods.

\subsection{Source Reliability Study}
\subsubsection{Method Comparison on Source Reliability Estimation}
The above experimental results demonstrate the superiority of the proposed method on both single truth finding and multiple claims ranking. In this experiment, we aim to study the precision of the estimated source reliability degrees. 

\begin{table}[h]
\setlength{\abovecaptionskip}{0pt}   
\setlength{\belowcaptionskip}{0pt}
\fontsize{7}{9}\selectfont
\centering
  \caption{Source Reliability Study on Synthetic Data}
  \begin{tabular}{cc}
    \toprule
    \multirow{2}{*}{Method} & Pearson's Correlation   \\
    &  (Reliability \& Error Rate) \\
    \midrule
    Investment & -0.9316\\
    PooledInvestment & -0.9216\\
    TruthFinder & -0.8858 \\
    CRH & -0.9816 \\
    WSDM 2017 & -0.9944\\
    The Proposed Method & \textbf{-0.9950} \\
  \bottomrule
\label{reliability_study}
\vspace{-0.5cm}
\end{tabular}
\end{table}

We adopt the Pearson's Correlation Coefficient between sources' error rates and their estimated reliability weights to measure the performance of source reliability estimation. As source reliability degrees and error rates are negatively correlated, the closer the Pearson Coefficient of the method is to -1, the better the method performs. Table~\ref{reliability_study} summarizes the results of all the methods. Compared with the baselines, the proposed method makes the best source reliability estimation. With a precise source reliability estimation, the proposed method can accurately determine the trustworthiness of the information provided by a source, which contributes to a better performance on the overall truth discovery task.

\subsubsection{Comparison of Source Reliability between Reference and Non-reference Sources}
\begin{figure}
\setlength{\abovecaptionskip}{0pt}   
\setlength{\belowcaptionskip}{0pt}
\centering
\includegraphics[width=0.32\textwidth]{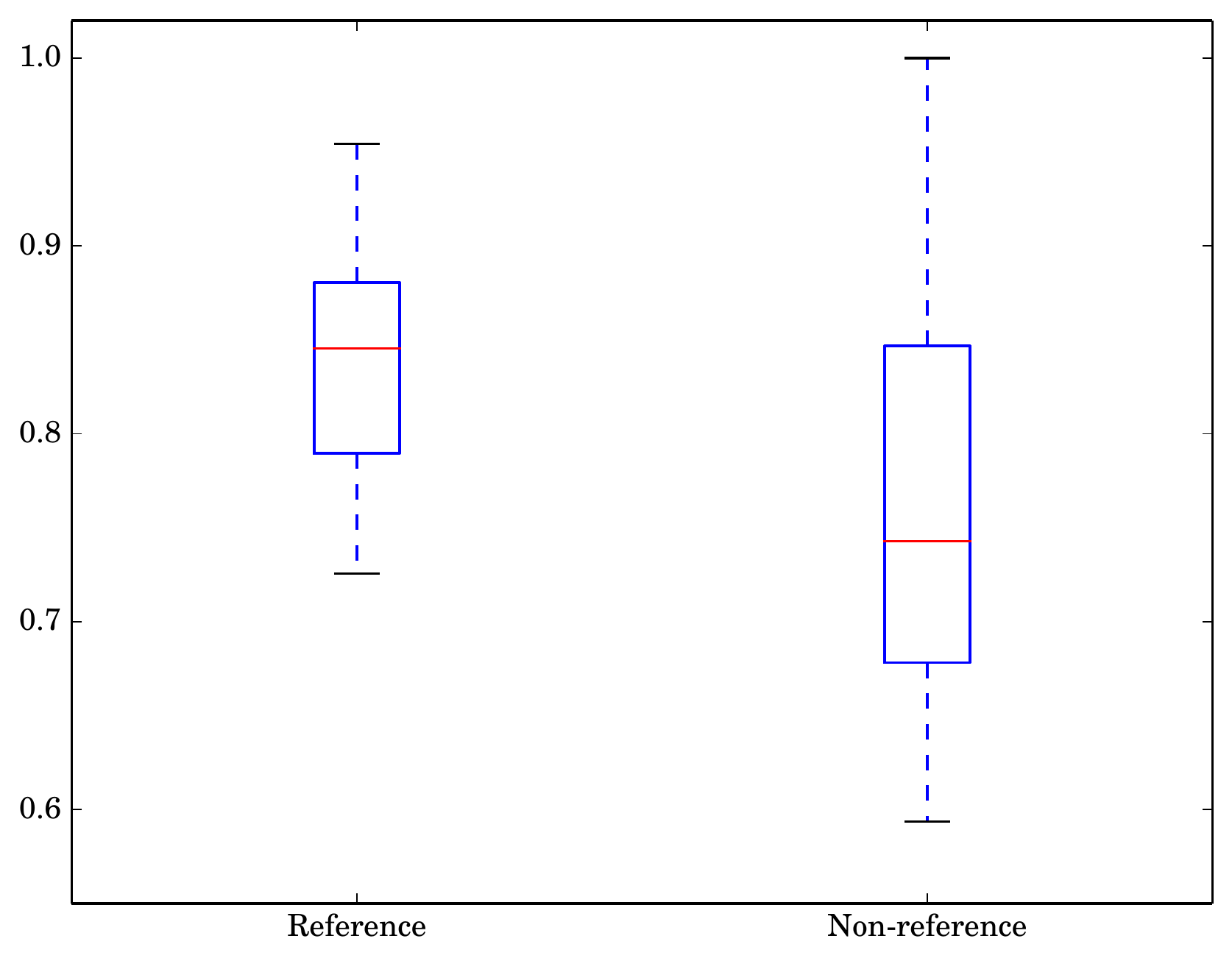}
\caption{Comparison between Ref. and Non-Ref. Sources}
\label{exp_5D}
\vspace{-0.5cm}
\end{figure}

As the proposed method aims to exploit priorly known high-quality reference sources to semi-supervise the truth discovery problem, we conduct an experiment to examine the difference in estimated source reliability degrees between reference and non-reference sources. We construct a synthetic dataset with 100 sources, 20 as reference sources, 80 as non-reference sources, in which we randomly add 0\%--50\% noise to generate reference sources from the ground truth data and add 0\%--100\% noise to generate non-reference sources. Fig.~\ref{exp_5D} shows that the overall estimated source reliability of reference sources is higher than that of non-reference sources. However, there are some non-reference sources that are estimated with a higher reliability than the reference sources. This result shows that the reference sources effectively semi-supervise the source reliability estimation to distinguish the reliable sources from the unreliable sources, instead of dominating the process.

\subsection{Effect of Integrating Object Embeddings}
Besides semi-supervised learning, the other motivation of the proposed method is to incorporate object embeddings to capture the correlation between similar objects. We design a simple but clear toy example to validate the motivation. As presented in Table~\ref{toy_exp}, there are three sources containing three objects and two claims.
\begin{table}[h]
\vspace{-0.4cm}
\setlength{\abovecaptionskip}{0pt}   
\setlength{\belowcaptionskip}{0pt}
\fontsize{7}{9}\selectfont
\centering
  \caption{Toy Example}
  \begin{tabular}{ccc}
    \toprule
    Source 1 &Source 2 & Source 3  \\
    \midrule
    Object 1 Claim 1 & Object 1 Claim 1& Object 1 Claim 1\\
    Object 2 Claim 1 & Object 3 Claim 2 & Object 1 Claim 2\\
  \bottomrule
\label{toy_exp}
\end{tabular}
\vspace{-0.5cm}
\end{table}

\begin{figure}
\setlength{\abovecaptionskip}{0pt}   
\setlength{\belowcaptionskip}{0pt}
\centering
\includegraphics[width=0.45\textwidth]{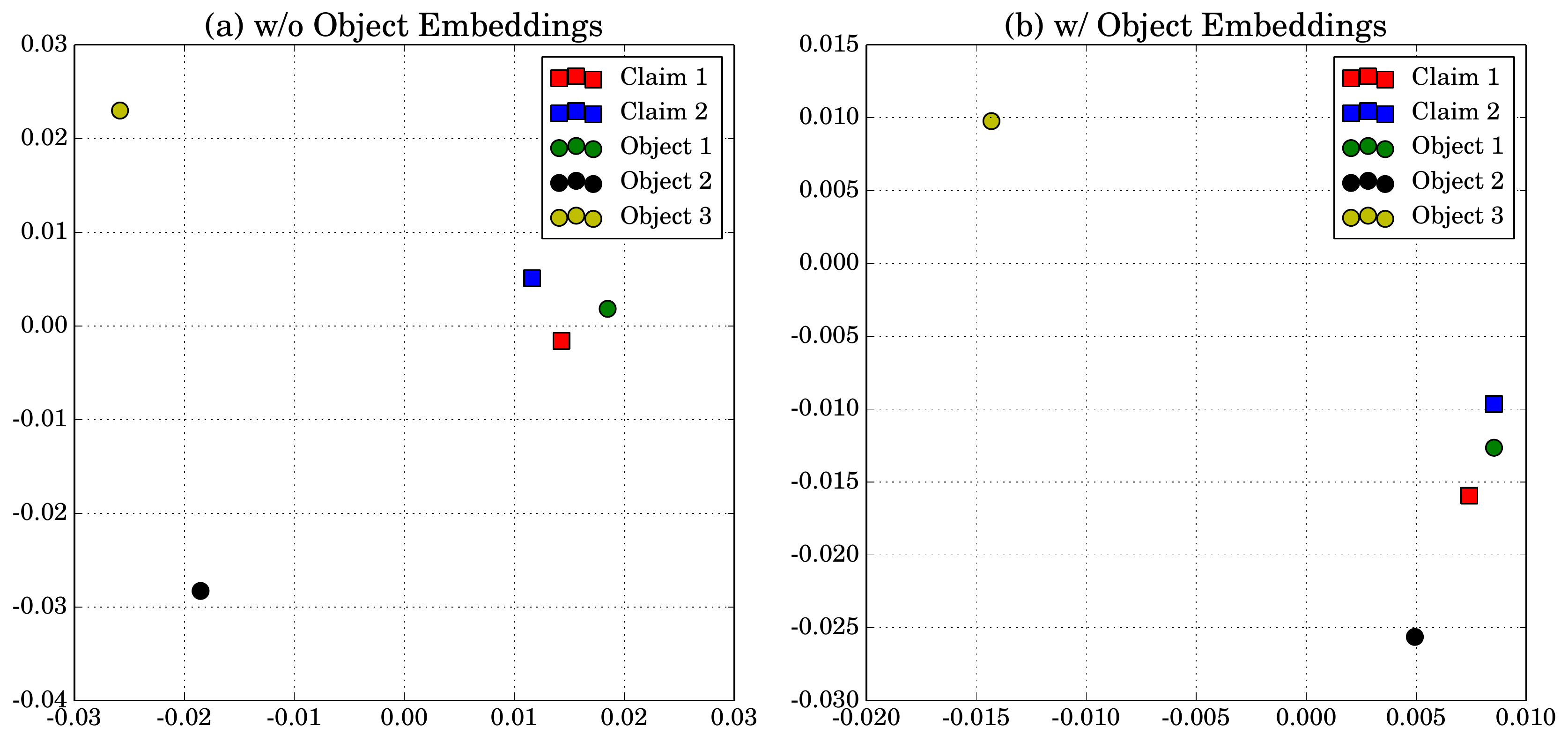}
\caption{Effect of Object Embeddings}
\label{exp_5E}
\vspace{-0.5cm}
\end{figure}

We assume that Object 1 and Object 2 have similar embeddings. We conduct an experiment to compare the proposed method with and without object embeddings. For without object embeddings, we use one-hot vector representations to represent each object, which means each object is independent. For with object embeddings, we assign a pair of similar embeddings for Object 1 and Object 2, while Object 3 is assigned with a relatively different embedding. We project the learned truth vectors and the condition representations onto a 2D plot in Fig.~\ref{exp_5E}. 

We can observe that the learned truth vectors and condition vectors are distributed independently when using one-hot vector representation for objects. But when using object embeddings, the learned truth vectors and condition vectors will be affected by the object embeddings so that objects with similar embeddings will learn similar truth vectors, so that they have similar distance with claims. For instance, Object 1 and Object 2 are relatively closer in Fig.~\ref{exp_5E}(b) than in Fig.~\ref{exp_5E}(a), since they are similar in embeddings. However, there is little difference between Object 2 and Object 3 in Fig.~\ref{exp_5E}(a), as the truth discovery process only takes into account the data itself, regardless of the correlation between objects.

\section{Experiment on Real-world Data}
The experiments on the synthetic datasets demonstrate the effectiveness of the proposed method under various truth discovery scenarios.

In this section, we conduct experiments on real-world medical datasets to validate the effectiveness of the proposed method from the following aspects:  (1) We compare the learned confidence scores and the statistic results to confirm that the proposed method can deal with the lack of high-quality data and discover knowledge condition from multi-source data. (2) In order to demonstrate the effectiveness of integrating the contextual embeddings of knowledge triples to enhance the correlation between similar triples, we compare different embedding methods and visualize the attention mechanism in the learning process.

\subsection{Data Description and Experimental Settings}
In this set of experiment, we focus on diseases that are related to \textit{chest pain}. Thus, we mainly collect the medical data pertaining to \textit{chest pain} and filter out those data that are not related to \textit{chest pain}.

We collect $336,670$ de-identified electronic medical records (EMRs) from hospitals. As this analysis is focusing on diseases that are related to chest pain, those records that are irrelevant and not containing any knowledge triple are filtered out. After preprocessing steps, we get $41,700$ EMR data from $360$ doctors. 
Then we collect question-answer pairs (QA pairs) from the largest medical QA website in China (http://www.xywy.com). After the same preprocessing steps as EMR data, we get $275,262$ QA pairs answered by $12,501$ users.

As the main focus of this paper is to discover the knowledge triple condition information, the knowledge triple extraction is regarded as the preprocessing step. To be more specific, we first utilize a constructed medical dictionary to extract medical entity mentions from the raw texts. Then we map these entity mentions to specific entity types. At last, we match entity pairs in the same text to possible knowledge triples with the help of a pre-constructed alias dictionary. By doing so, we extract the knowledge triple mentions from the raw medical data.

During the discussion of experimental results, we abbreviate the expression of a knowledge triple to an entity pair, (e.g., we abbreviate \textit{(coronary heart disease, disease-symptom, chest pain)} to \textit{(coronary heart disease, chest pain)}), as the original knowledge triple can be very long.

\subsection{Condition Confidence Score}
We first make a statistical analysis on the frequency distribution of knowledge triple occurrence on both the EMR dataset and the QA dataset to show the insufficiency of the EMR data and the necessity of incorporating external data. Then we sample some typical cases to study the ability of the proposed method on discovering knowledge condition with limited reliable high-quality data. These typical cases include knowledge triples that rarely appear in EMR but appear frequently in QA, knowledge triples that rarely appear in both EMR and QA, and knowledge triples that appear frequently in both EMR and QA.
  
\subsubsection{Statistics of the Frequency Distribution of Knowledge Triple Occurrence}
\begin{figure}
\setlength{\abovecaptionskip}{0pt}   
\setlength{\belowcaptionskip}{0pt}
\centering
\includegraphics[width=0.4\textwidth]{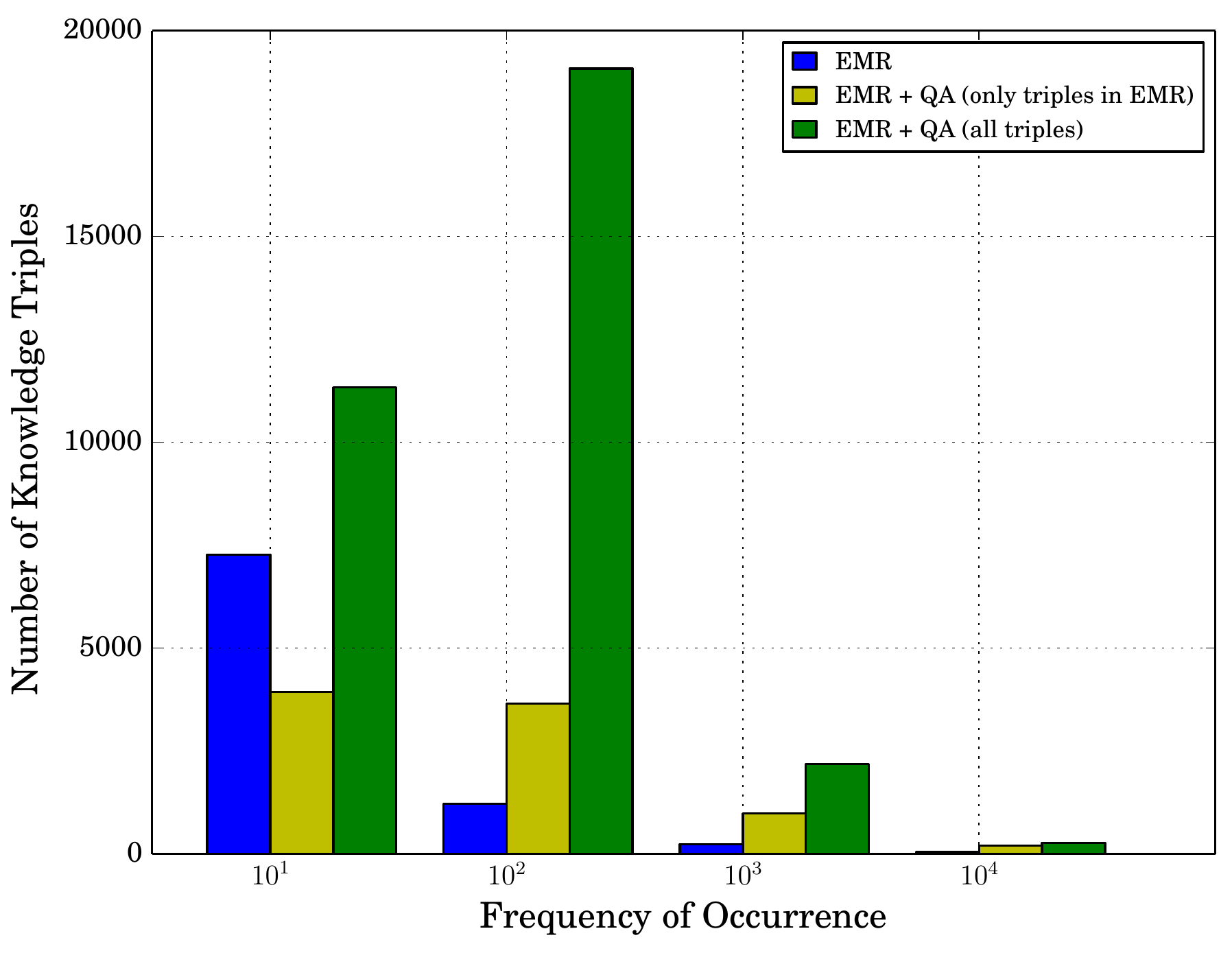}
\caption{Frequency Distribution of Triple Occurrence}
\label{stat}
\vspace{-0.5cm}
\end{figure}

Fig.~\ref{stat} shows the frequency distribution of knowledge triple occurrence. Compared with only using the EMR data, incorporating QA data effectively increases the frequency of triple occurrence, which also provides more condition information. Meanwhile, QA data also brings a large number of new knowledge triples, enabling us to obtain condition information of more triples.

\subsubsection{Case Study -- Knowledge triples that rarely appear in EMR but appear frequently in QA}
Table~\ref{case1} shows the count results and the truth discovery results of knowledge triple \textit{(breast hyperplasia, chest pain)} and condition \textit{(Gender)}. Following the count result on EMR data, we may come to the conclusion that the fact \textit{(breast hyperplasia, chest pain)} has only one gender condition, \textit{female}. Actually, the knowledge triple \textit{(breast hyperplasia, chest pain)} also appears in men, which is proved by the QA data. Thus, for most of knowledge triples that rarely occur in EMR data, we can obtain a more precise condition information by incorporating QA data.

\begin{table}[h]
\vspace{-0.3cm}
\setlength{\abovecaptionskip}{0pt}   
\setlength{\belowcaptionskip}{0pt}
\fontsize{7}{9}\selectfont
\centering
  \caption{Case Study of \textbf{(breast hyperplasia, chest pain)}.}
  \begin{tabular}{cccc}
    \toprule
    \multicolumn{4}{c}{(breast hyperplasia, chest pain)}\\
    \midrule
    Condition  & EMR  & QA   & Confidence  \\
    (Gender) &Count&Count & Score\\
    \midrule
    female &1 & 879& 1.00\\
    male &0 & 112 &0.77\\
      \bottomrule
\label{case1}
\vspace{-0.6cm}
\end{tabular}
\end{table}

\subsubsection{Case Study -- Knowledge triples that rarely appear in both EMR and QA} 
Despite the large amount of condition information provided by the additional QA data, there are still a small number of knowledge triples that rarely appear in both EMR and QA data. Table~\ref{case2} summarizes the count results and the learned condition confidence scores of knowledge triple \textit{(mycoplasma pneumoniae pneumonia, chest pain)}. The count results in both EMR and QA data show that only very few data contains this knowledge triple. However, we can observe from the learned condition confidence scores that there is still a degree of differentiation between the conditions, from which we can learn that the knowledge triple \textit{(mycoplasma pneumoniae pneumonia, chest pain)} has little difference in gender condition but inclines to the young in age condition. 
The knowledge triple \textit{(pneumonia, chest pain)}  is the most similar triple with \textit{(mycoplasma pneumoniae pneumonia, chest pain)} in the learned truth vector space. (The experimental results of the learned truth vectors will be discussed later.) We can see that these two knowledge triples share a similar ranking of related conditions, which indicates that the proposed method enables the rare knowledge triples to learn the condition information from similar triples.

\begin{table}
\setlength{\abovecaptionskip}{0pt}   
\setlength{\belowcaptionskip}{0pt}
\fontsize{7}{9}\selectfont
\centering
  \caption{Case Study of \textbf{(mycoplasma pneumoniae pneumonia, chest pain)}.}
  \begin{tabular}{cccccc}
    \toprule
    \multicolumn{4}{c}{(mycoplasma pneumoniae pneumonia, chest pain)}&\multicolumn{2}{c}{(pneumonia, chest pain)}\\
    \cmidrule(lr){1-4} \cmidrule(lr){5-6}
    Condition  & EMR  & QA  & Confidence &All & Confidence \\
    (Gender\&Age) &Count&Count & Score&Count & Score\\
    \midrule
    male &0 & 4 &1.00&1698&1.00\\
    female &1 & 1&0.92&991&0.91\\
    \midrule
    20 &1 & 3&1.00&815&0.99\\
    10 &0 & 0 &0.96&255&1.00\\
    30 &0 & 1&0.87&425&0.86\\
    0 &0 & 0 &0.81&267&0.86\\
    40 &0 & 1&0.67&329&0.69\\
    90 &0 & 0 &0.54&19&0.55\\
    50 &0 & 0&0.54&262&0.58\\
    60 &0 & 0 &0.51&200&0.54\\
    80 &0 & 0&0.49&25&0.50\\
    70 &0 & 0 &0.48&79&0.50\\
      \bottomrule
\label{case2}
\vspace{-0.8cm}
\end{tabular}
\end{table}

\subsubsection{Case Study -- Knowledge triples that appear frequently in both EMR and QA} 
Table~\ref{case4} summarizes the results of two common knowledge triples in both EMR and QA data, \textit{(coronary heart disease, chest pain)} and  \textit{(bronchitis, chest pain)}, and their corresponding age conditions. Compared with the statistic probability results, the learned confidence scores have a similar distribution over all the age condition but indicate a more obvious condition inclination. The triple \textit{(coronary heart disease, chest pain)} is more related to the old, while \textit{(bronchitis, chest pain)} inclines to the young. In order to testify the effectiveness of the semi-supervision, we conduct ablation test of disabling the semi-supervision from reference sources, in which both EMR and QA data are regarded as non-reference sources. As shown in Table~\ref{case4}, in both cases, the distributions of confidence score are more discriminative with semi-supervision, as unreliable information is assigned with more punishment while reliable information is awarded with higher trustworthiness. This phenomenon justifies that the proposed method successfully exploit limited reliable medical data, to semi-supervise the task of medical knowledge condition discovery.

\begin{table*}[thpb]
\setlength{\abovecaptionskip}{0pt}   
\setlength{\belowcaptionskip}{0pt}
\fontsize{7}{9}\selectfont
\centering
  \caption{Case Studies of \textbf{(coronary heart disease, chest pain) and \textbf{(bronchitis, chest pain)}}}
  \begin{tabular}{cccccccccc}
    \toprule
    \multicolumn{5}{c}{(coronary heart disease, chest pain)}&\multicolumn{5}{c}{(bronchitis, chest pain)}\\
    \cmidrule(lr){1-5} \cmidrule(lr){6-10}
    Condition &All & \multicolumn{2}{c}{Confidence Score} & Statistic&Condition &All & \multicolumn{2}{c}{Confidence Score}& Statistic \\
    (Age)&Count & w/ ref & w/o ref& Probability& (Age)&Count&  w/ ref & w/o ref & Probability\\
    \midrule
    60 &1146&1.00&1.00&18.42\%&0 & 116 &1.00&1.00&6.66\%\\
    50 &1416&0.98&0.99&22.75\%&10 &196 &0.99&0.99&11.24\%\\
    70 &556&0.94&0.95&8.93\%&20 &591  &0.95&0.96&33.91\%\\
    80 &128&0.88 &0.90&2.06\%&30  & 367&0.87&0.90&21.06\%\\
    40 &1253&0.86 &0.87&20.14\%&40 & 233 &0.71&0.79&13.37\%\\
    90 &8&0.75 &0.80&0.13\%&50 &134 &0.59&0.66&7.69\%\\
    30 &770&0.69&0.76&12.37\%&90 & 3 &0.58&0.66&1.72\%\\
    10 &77&0.66&0.73&1.24\%&60  &59 &0.55&0.65&3.38\%\\
    20 &659&0.64&0.68&10.59\%&70 &39  &0.51&0.59&2.24\%\\
    0 &210&0.45&0.52&3.37\%&80 & 5 &0.51&0.62&2.87\%\\
  \bottomrule
\label{case4}
\vspace{-0.5cm}
\end{tabular}
\end{table*}

\subsection{Case Study of the Learned Truth Vector for Knowledge Triples}

\subsubsection{Comparison of Knowledge Triple Embeddings Methods} In order to capture the correlation between different knowledge triples, we incorporate knowledge triple embeddings into the proposed method. We combine the co-occurrence embeddings learned from CBOW and the head and tail entity embeddings in the knowledge triple to represent the knowledge triple. In this experiment, we compare this knowledge triple embedding method with two baselines by a case study that ranks the distance between learned truth vectors $v_m^*$ to observe the similarity between knowledge triples.

From Table~\ref{case_truth} we can observe that: 1) In the case of only entity embeddings, all the top-ranked similar triples contain those entities most similar to \textit{coronary heart disease} or \textit{chest pain}, such as \textit{heart disease} and \textit{left chest pain}. 2) With only co-occurrence embeddings, the top-ranked similar triples are diverse but somehow related to \textit{(coronary heart disease, chest pain)}, like \textit{(cardio-cerebrovascular disease, chest pain)} and \textit{(myocardial infarction, filling defect)}, since they often occur in the same medical case. 3) When combining these two kinds of embeddings, we can obtain a comprehensive result that takes into account not only the similarity in entity level but also the co-occurrence information in triple level.

\begin{table*}
\setlength{\abovecaptionskip}{0pt}   
\setlength{\belowcaptionskip}{0pt}
\fontsize{7}{9}\selectfont 
\centering
  \caption{The triples that learn similar truth vectors to \textbf{(coronary heart disease, chest pain)} with different embeddings.}
  \begin{tabular}{cccccc}
    \toprule
    \multicolumn{2}{c}{Only Entity Embeddings} & \multicolumn{2}{c}{Only Co-occurrence Embeddings} & \multicolumn{2}{c}{Combine Two Kinds of Embeddings}\\
    \cmidrule(lr){1-2} \cmidrule(lr){3-4} \cmidrule(lr){5-6}
    Similar Triple  & Distance &Similar Triple  & Distance & Similar Triple & Distance \\
    \midrule
   (heart disease, chest pain) & 0.9872 & (heart disease, chest pain) & 0.6546 &(heart disease, chest pain) & 0.9444 \\
   (myocardial infarction, chest pain) & 1.0204 & (cardio-cerebrovascular disease, chest pain) & 0.7095 &(cardio-cerebrovascular disease, chest pain) & 1.0933 \\
   (coronary heart disease, left chest pain) & 1.2282 & (cardiovascular disease, coronary insufficiency) & 0.7275 &(coronary heart disease, dorsal distending pain) & 1.2294 \\
   (coronary heart disease, chest distress) & 1.2822 & (myocardial infarction, filling defect) & 0.7504 &(heart disease, limited activity) & 1.2520 \\
   (heart failure, chest pain) & 1.4299 & (coronary heart disease, shoulder pain) & 0.8885 & (myocardial infarction, filling defect) & 1.2703 \\
  \bottomrule
\label{case_truth}
\vspace{-0.6cm}
\end{tabular}
\end{table*}

\subsubsection{Effect of Attention Mechanism}
The proposed method provides an intuitive way to inspect the importance degree among all the triples in the same case by visualizing the attention weight $\alpha_i$ from Eqn. (3). Due to the limited space, we randomly choose one case from the dataset and visualize the attention weights.
\begin{figure}
\setlength{\abovecaptionskip}{0pt}   
\setlength{\belowcaptionskip}{0pt}
\centering
\includegraphics[width=0.45\textwidth]{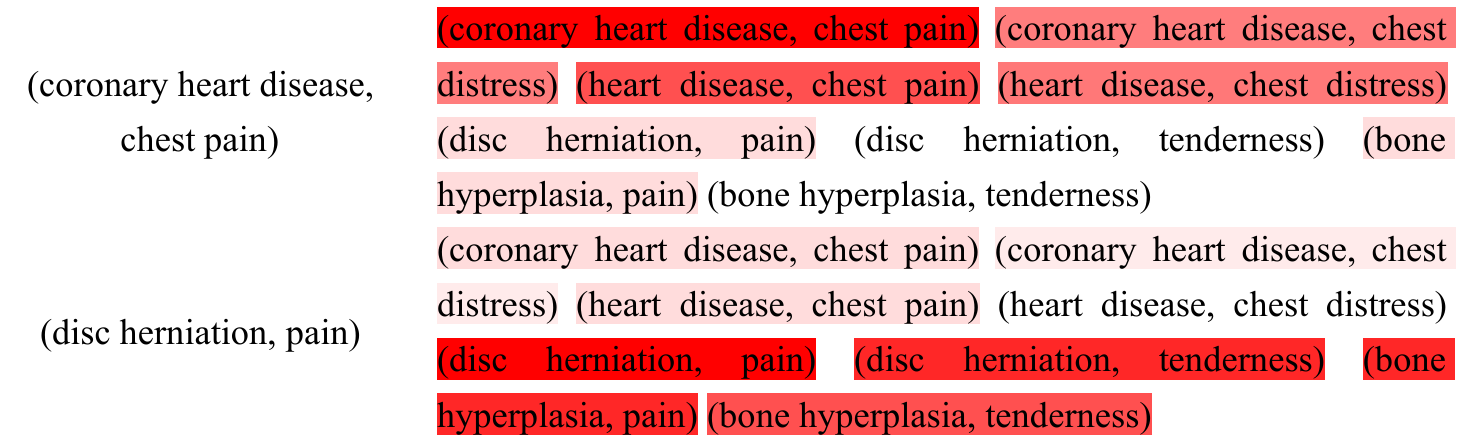}
\caption{Visualization of Attention Weights}
\label{attention}
\vspace{-0.6cm}
\end{figure}

In Fig.~\ref{attention}, the color depth indicates the contribution degree of the triples in the context, the darker the color, the more the contribution. When calculating the contextual embeddings for \textit{(coronary heart disease, chest pain)}, those related triples make a greater contribution, such as \textit{(coronary heart disease, chest distress)} and \textit{(heart disease, chest pain)}. In contrast, for the triple \textit{(disc herniation, pain)}, \textit{(disc herniation, tenderness)} and \textit{(bone hyperplasia, pain)} contribute more.
\section{Related Work}
\textbf{Knowledge Discovery from EMRs}.
With the proliferation of Electronic Medical Records (EMRs), such high-quality medical data open a new window for data-driven knowledge discovery towards medical decision support. There are various medical knowledge discovery applications based on EMRs, including medical entity discovery~\cite{Jochim2017Named}, disease topic discovery~\cite{Li2016Hierarchical}, temporal pattern mining~\cite{Tourille2017Neural}, medical event detection~\cite{Jagannatha2016Bidirectional}. In this paper, we study medical knowledge condition information discovery from EMRs, which plays a crucial role in medical systems and related applications.

\noindent \textbf{Enriching Knowledge Graph}.
Nowadays, several large-scale knowledge graphs have been constructed, such as DBpedia~\cite{Lehmann2015DBpedia}, YAGO~\cite{Hoffart2011YAGO2}. However, these knowledge graphs are often incomplete or uncertain due to the limitation of available human knowledge or the probabilistic nature of real-world knowledge. Thus, it is often desirable to supplement the knowledge graph with extra information, such as type information~\cite{Moon2017Learning}, temporal information~\cite{Hoffart2011YAGO2}, probability information~\cite{Chen2014Knowledge}. The way to enrich the knowledge graph can be divided into two groups.  The first is to enrich the distributed knowledge representation by incorporating extra knowledge into knowledge embeddings~\cite{Moon2017Learning,Xie2017Does}. The other way is to reconstruct the knowledge graph with some new elements, such as probability~\cite{Jiang2017Learning,Chen2014Knowledge}, time and space constraint~\cite{Hoffart2011YAGO2,Chekol2017Marrying}. Following the second fashion, we focus on the condition information in the knowledge graph, which has received little attention so far.

\noindent \textbf{Truth Discovery}.
Truth discovery aims to estimate the reliability of different sources to integrate multi-source noisy information~\cite{Li2016A}. In the last decade, many truth discovery methods~\cite{Yin2011Semi,Liu2011Online,Dong2012Less,Li2014Resolving,Pasternack2010Knowing,Yin2007Truth} have been proposed to estimate source reliability degrees and identify trustworthy information. Among these truth discovery methods, some utilize a subset of labeled truth to semi-supervise the process of source reliability estimation and truth computation~\cite{Yin2011Semi,Liu2011Online,Dong2012Less}. Different from these methods, we adopt priorly known high-quality reference sources to semi-supervise the overall truth discovery process. 
Most of the current truth discovery methods~\cite{Li2014Resolving,Pasternack2010Knowing,Yin2007Truth} compute the truth for each object independently, or take into account the interaction between claims~\cite{li2017reliable} and sources~\cite{Lyu2017Truth}. Different from the aforementioned methods, the proposed truth discovery method encodes the object information to capture the correlation between objects in truth discovery task.

Recently, several efforts have been made on applying truth discovery to real-world applications, such as knowledge base construction~\cite{Dong2014Knowledge,DBLP:conf/sigmod/XinMC18}, question answering~\cite{li2017reliable,Ma2015FaitCrowd}, information retrieval~\cite{Li2011T,Li2015Truth}. Especially in medical and health domain, truth discovery methods~\cite{li2017reliable,DBLP:conf/kdd/ZhangLDFY18,DBLP:journals/tiis/DumitracheAW18} can effectively deal with large amount of noisy multi-source medical data with diverse quality. In this paper, we design a new truth discovery method to discover the knowledge condition information from noisy multi-source medical data.

\section{Conclusions}
In this paper, we present a medical knowledge condition discovery method to enrich medical knowledge graph with condition information. Due to the limited amount of available EMR data, we leverage medical QA data from online crowdsourcing medical communities to overcome the lack of data. However, unlike EMR data, the quality of QA data is diverse, as the answers are provided by website users with different professional levels, which may introduce a lot of noise and degrade the quality of discovered conditions. To tackle these challenges, we propose a novel truth discovery method for the task of medical knowledge condition discovery. The proposed method can recognize the EMR data as priorly known high-quality reference sources to semi-supervise the overall process of medical knowledge condition discovery in multi-source medical data. Besides, the proposed method incorporates the occurrence and entity information of knowledge triples to capture the interaction between knowledge triples when computing the truth for knowledge triples. Experimental results on real-world medical datasets show that the proposed method can effectively discover accurate medical knowledge condition information from multi-source data with diverse quality. We also validate the effectiveness of the proposed method under various scenarios on synthetic datasets. 
\label{sec:conclusion}

\section{Acknowledgments}
This work was financially supported by the National Natural Science Foundation of China (No.61602013), the Shenzhen Fundamental Research Project (No.JCYJ20170818091546869), the Shenzhen Project (No.ZDSYS201802051831427), and the project "PCL Future Regional Network Facilities for Large-scale Experiments and Applications (PCL2018KP001)". Min Yang was sponsored by CCF-Tencent Open Research Fund.

\balance
\bibliographystyle{ACM-Reference-Format}
\bibliography{bibliography}

\end{document}